\documentclass[reprint,aip,superscriptaddress,floatfix,showpacs]{revtex4-1}
\usepackage[utf8]{inputenc}
\usepackage{amsmath}
\usepackage{amssymb}
\usepackage{graphicx}

\makeatletter
\usepackage{graphics}
\usepackage{epsfig}
\usepackage{color}

\newcommand{\be}{\begin{equation}} \newcommand{\ee}{\end{equation}}
\newcommand{\bea}{\begin{eqnarray}} \newcommand{\eea}{\end{eqnarray}}
\graphicspath{ {Fig/} }

\bibpunct{[}{]}{,}{n}{}{}
\renewcommand{\onlinecite}[1]{\cite{#1}}

\makeatother

\begin{document}

\title{A mean field approach to multiple, long-delayed systems}

\author{Giovanni Giacomelli}
\affiliation{Consiglio Nazionale delle Ricerche, Istituto dei Sistemi Complessi, via Madonna del Piano 10, I-50019 Sesto Fiorentino (FI), Italy}

\author{Antonio Politi}
\affiliation{Consiglio Nazionale delle Ricerche, Istituto dei Sistemi Complessi, via Madonna del Piano 10, I-50019 Sesto Fiorentino (FI), Italy}
\affiliation{Institute of Pure and Applied Mathematics, Department of Physics (SUPA), Old Aberdeen, Aberdeen AB24 3UE, United Kingdom}

\date{\today}

\begin{abstract}
The concept of multiple, long-delayed feedback systems is introduced and discussed with reference to a paradigmatic model.
We analyse how the resulting chaotic dynamics is affected by the delay distribution. Via a mean-field approach, we show that
a spatio-temporal representation equivalent to the one developed for the single-delay can be extended to this wider class of dynamical systems.
Numerical simulations are complemented by a theoretical study based on a multiple-scale analysis, which, in the vicinity of a Hopf bifurcation, allows mapping the initial model onto a complex Ginzburg Landau equation.
As a result, we find that the only relevant feature influenced by the multiple delays is the size of the coherent spatio-temporal structures which, in turn, depends exclusively on a generalized {\it variance} of the delay distribution.

\end{abstract}

\maketitle

{\bf 
Lattice structures are widely studied in many fields, such as graphene and carbon nanotubes~\cite{katsnelson2012graphene}, the Ising model~\cite{onsager1944crystal}, percolation theory~\cite{stauffer2018introduction}, self-avoiding walks~\cite{madras2013self}, up to lattice QCD~\cite{wilson1974confinement} and cellular automata~\cite{gardner1970mathematical}.
In the case of macroscopic lattices, such as optical networks~\cite{Vynck2023}, the propagation times are large with respect to the typical system time-scales, suggesting the presence of a long-delay regime~\cite{Yanchuk2017}. In this Letter, we introduce and discuss a broad framework where several long delayed feedbacks are simultaneously active, all with the same nominal lag: as a paradigm of real systems, the delay times fluctuate around an average value. We discuss a simple but general model, and compare numerical simulations with a normal form obtained via a multiscale analysis, showing that the dynamics is essentially determined 
by the width of the delay distribution. As a consequence, we show that the system is strongly resilient to perturbations of the microscopic parameters (i.e., delay times and couplings). 
}

\subsection{Introduction}
In real physical systems, finite-propagation velocities always lead to delays in the transmission of information. This is a particularly relevant feature of
 the mammalian brain, where the delay is indeed due to the propagation time of action potentials along the axons~\cite{Campbell2007}.
 In many cases, this effect strongly affects the overall dynamics, possibly increasing the dynamical complexity of the stationary regime~\cite{Farmer1982, Chaos2017}. 
Delayed interactions have been investigated in several physical contexts: from the control and stabilization of the asymptotic regime~\cite{Flunkert2013}, to the implementation of the reservoir computing protocol~\cite{Illing2019}.
A particularly relevant class of systems is that where the delay is much longer than the intrinsic timescales of the system without feedback (for a review, see e.g. \cite{Yanchuk2017}). This setup naturally induces a multiscale regime, which can be effectively described by introducing new independent variables:
a pair of spatial and temporal coordinates~\cite{Arecchi1992}. 
This  parametrization goes well beyond a formal description; it allows developing a mathematical formalism, able to grasp and reframe the overall scenario.
The most peculiar feature is the structure of the temporal autocorrelation, with sharp revivals which evidence the presence of a multiscale dynamics.
 They are located at multiples of $T+v$, where $T$ is the delay and $v\simeq 1$ is typically a small correction. In such a case, the spatio-temporal representation can be carried out by decomposing the time variable as 
$t= \sigma +\theta T$, where $\sigma \in [0,T]$ is a pseudo-spatial variable ($T$ plays the role of the system size) and $\theta$ is an (integer) pseudo-temporal variable. The temporal series can be accordingly re-organized as spatio-temporal patterns where otherwise hidden, propagating structures (characterized by a drift $v$) become clearly visible. In some contexts, the field variable(s) change very slightly from one to the next delay unit, allowing for the representation of the evolution rule via a partial differential equation where $\theta$ can be treated as a continuous variable. This happens, for instance, in the vicinity of a Hopf bifurcation, as thoroughly discussed in \cite{Giacomelli1996}. Another interesting example is laser dynamics, where the almost continuity along the time axis is a consequence of the vicinity to a perfectly integrable limit~\cite{Politi2023}.

 So far, most of the literature focused on systems characterized by a single delay. An exception is that of neural fields, where distribution of delays are naturally included,  with the goal of, e.g., identifying the relevant bifurcations as in Ref.~\cite{Atay2004}. On a more general level, multiple, hierarchically long delayed systems have been considered in Ref.~\cite{Yanchuk2014-2015}. Here we consider a setup where multiple long similar delays are simultaneously present, showing that in a first approximation the
 dynamics is qualitatively equivalent to that of an equivalent single-delay model (mean field description).
 More precisely, we find that the resulting stationary regime is determined by the dispersion of the delays, quantified by a generalized variance, where the term
 {\it generalized} means that the quantity is determined by averaging over positive as well as negative coupling strengths (akin to the excitatory and inhibitory synapses 
 present in neuronal networks~\cite{vanVreeswijk1996,vanVreeswijk1998}).
 
 In Sec.~\ref{sec:numerics}, we present a series on numerical results to illustrate the dynamics for various distributions of feedback distributions and different points of views.
 In Sec.~\ref{sec:theory}, we develop two theoretical approaches: first we perform a linear stability analysis of the stationary fixed-point solution to show that, at leading
 order, the transition point is not affected by the presence of multiple delays. Then, by implementing a third-order multiple-scale perturbative analysis, we show that
 the dynamics of the delayed system is equivalent to that of a Complex Ginzburg-Landau (CGL) equation, similarly to the single delay model, but with the
 important difference that the diffusione coefficient depends also on the dispersion of the delays.
 Finally, open problems and possible generalizations are discussed in Sec.~\ref{sec:conclusion}.

\subsection{The model and numerical results}\label{sec:numerics}
We address the problem by studying a simple model, 
the multi-delay Stuart-Landau equation,
\begin{equation}
x_t = -x - (1+i\beta) |x|^2x +  \sum_j a_j x(t-T_j)   \;.
\label{eq:model}
\end{equation}
The first two terms in the r.h.s. account for the local dynamics, while the last one 
accounts for a feedback realized by a linear combination of processes with different delays $T_j$. It is convenient to express the feedback term by referring to a global coupling strength $\eta=\sum_j a_j$ and introducing
normalized weights $w_j = a_j/\sum_j a_j $. It is also convenient to write $T_j=T+\delta_j$, where $T=\sum_j w_j T_j =\langle T_j\rangle$ denotes the average delay, and $\delta_j$ is the deviation characterizing each single process with $\langle \delta_j\rangle=0$. 
Altogether, the ensemble of delays is described by the distribution $W(\delta)$, which is not necessarily positive defined, since the single feedbacks
can in general be either positive or negative. It is only important to keep in mind that nontrivial oscillations may emerge only when excitatory coupling
prevails: this is implicit in the assumption that $\int d\delta  W(\delta) =1$. Since all averages are weighted according to the coupling stength, 
they generally differ from the naive arithmetic mean.

The model (\ref{eq:model}) is the simplest description of the dynamics of a Stuart-Landau oscillator \cite{Landau1944,Stuart1960} on a macroscopic lattice \cite{Yanchuk2012,Larger2012,Brunner2013}, i.e. systems where the propagation delays along the links are larger than the typical timescale of the single oscillator.  
As such, it represents a generalization of the single-delay case considered in Ref.~\onlinecite{Giacomelli1996}.
So long as the delay dispersion around $T$ is small, we expect the overall scenario to be qualitatively similar.
In particular, we expect the Hopf bifurcation discussed in Ref.~\cite{Giacomelli1996} to be still 
present in (\ref{eq:model}) with an associated emergence of long time scales close to the bifurcation point. 
Hence, we start performing some numerical simulations of (\ref{eq:model}) for $\beta=3$ and $\eta=1.03$ (i.e, slightly above the threshold
value $\eta=1$ found in the single delay case). We first assume a flat distribution of delays around a large $T$, for different values of the delays dispersion as quantified  by the variance $M_2 = \langle \delta_j^2\rangle$. We observe in all the cases an irregular behavior, with strong intensity oscillations. In order to make the analysis more quantitative, we study the normalized autocorrelation function of the field amplitude $X(t)=\sqrt{xx^*}$,
\[
C(s) = \frac{\overline{X(t)X(t-s)}-\overline{X(t)}^2}{\overline{X^2(t)}-\overline{X(t)}^2}
\]
where the overbar denotes a time average.

\begin{figure}
\includegraphics[width=0.9\linewidth]{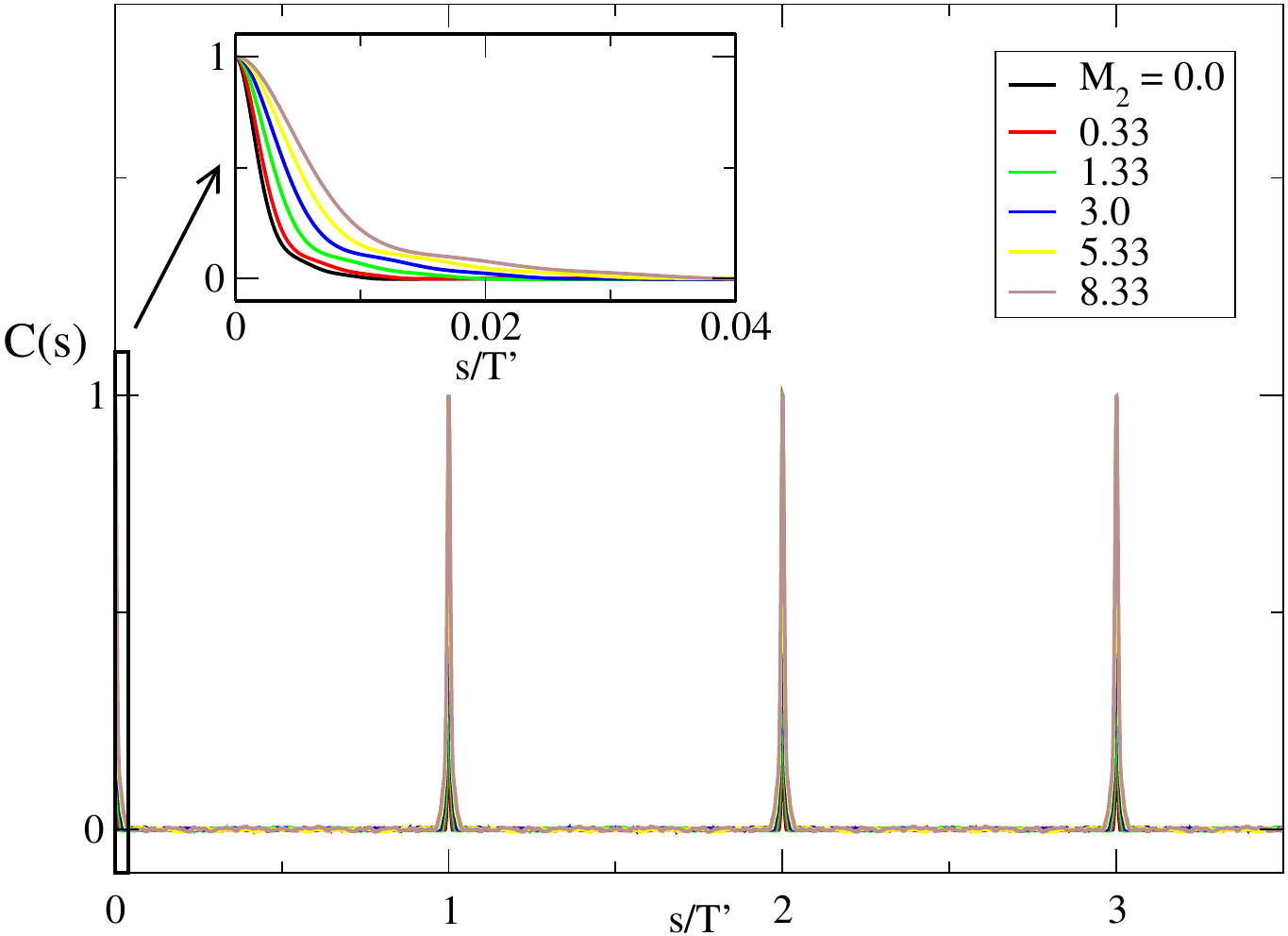}
\caption{Autocorrelation function for $\eta = 1.03$, using an uniform distributions of delays within the interval $[T-L/2,T+L/2]$ for $T=2048$ and $L=0,2,4,6,8,10$. The values reported in the legend are the corresponding variance $M_2=L^2/12$. Inset: zoom of initial part. Here, $T'=T+1$.}
\label{fig1}
\end{figure}

In Fig.~\ref{fig1} we present the results obtained from the numerical simulations. The autocorrelations display an initial fast decay - a signature of a chaotic regime -  accompanied by sharp revivals at multiples of $T'=T+1$. Such a behavior, found for different choices of large $T$ values is indeed analogous to what observed in the single delay case \cite{Arecchi1992} and suggests a possible equivalence with a standard long-delay description \cite{Yanchuk2017} with reference to the average delay $T$. The correspondence is strengthened by looking at the position of the peaks which is unaltered in the presence of multiple delays.
On the other hand, the inset  of Fig.~\ref{fig1}, where an enlargement of the first peak is displayed, shows that the
width of the peaks increases upon broadening the distribution of delays (here quantified by $M_2$), indicating a first effect of the feedback multiplicity.

\begin{figure}[htbp]
\makebox[0pt][c]{\includegraphics[width=1.1\columnwidth]{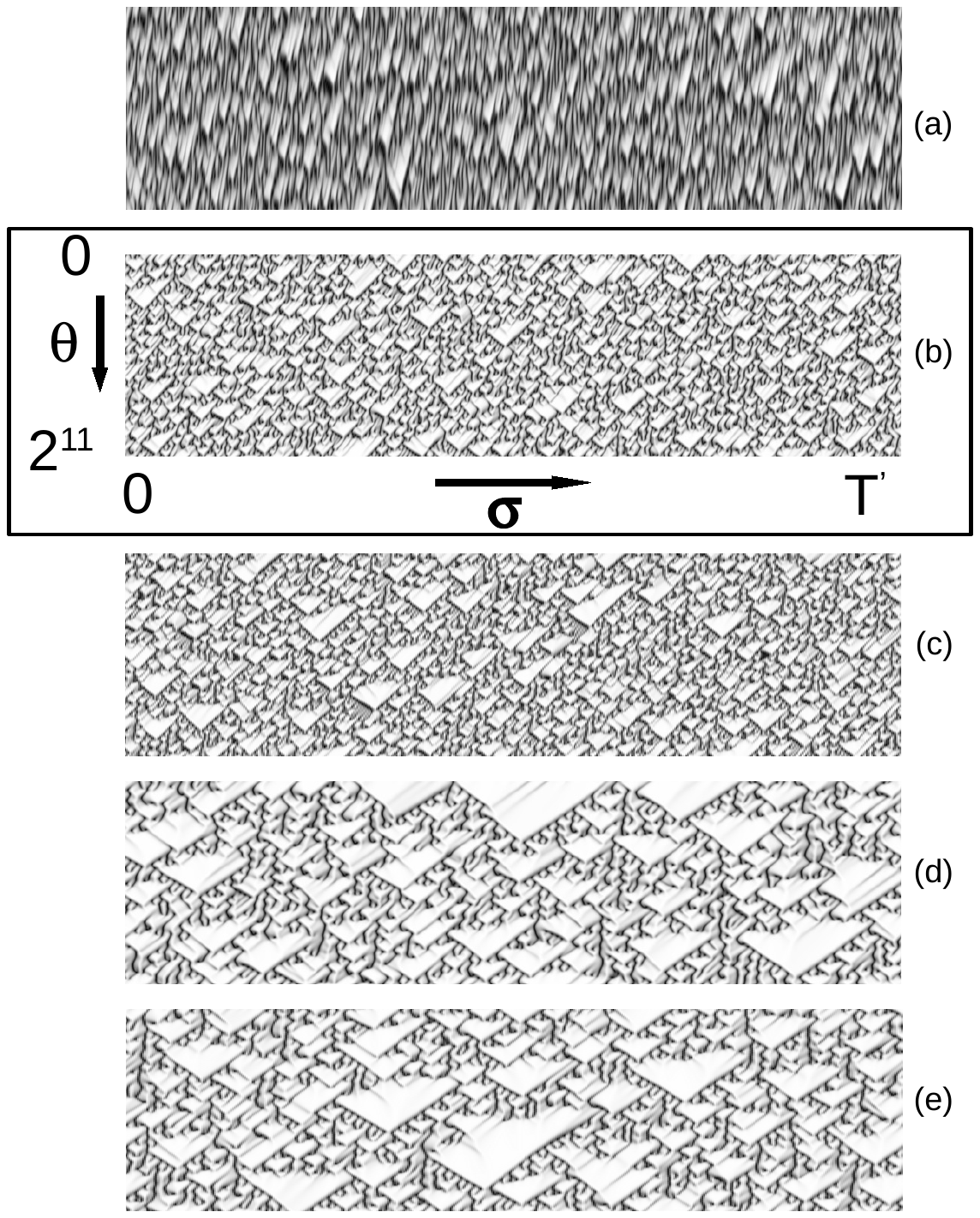}}
\caption{Patterns obtained for $\eta=1.10$ and $T=2048$: 
(a) two delays $T_1=2048.4$ and $T_2=2048.8$, with  $w_1=2$ and $w_2=-1$ ($M_2=-0.32$); (b) (boxed): single delay; 
(c) quasi-distribution of delays (see text) with $M_2=0$;
(d) uniform distribution with $M_2=1.33$;
(e) two delays $T_1=2049.5$ and $T_2=2047.1$, with $w_1=0.372$ and $w_2=0.698$ ($M_2=1.33$).
The drift is removed by adjusting the spatial length by $v\simeq1$ unit (i.e. $T'=T+1$), and the scales are as in (b).}
 \label{fig2}
\end{figure}

More detailed information can be obtained from the study of the spatio-temporal representations reported in Fig.~\ref{fig2}. The pattern for the single-delay case is plotted in the box (Fig.~\ref{fig2}b) as a reference, while
Fig.~\ref{fig2}d refers to a flat distribution of delays with $M_2=1.33$. It is seen that the dynamical regime is almost the same as for the single delay case, which was previously identified as an instance of chaotic dynamics within the Benjamin-Feir stable region\cite{Benjamin1967, Lange1974} of the equivalent CGL model~\cite{Chate1994,Aranson2002}. The only apparent difference is the increase of the spatial scale of the coherent structures: a visual manifestation of the slower decay of the autocorrelation visible in Fig.~\ref{fig1}.

In an effort to empirically identify those features of the weight distribution actually responsible for the asymptotic pattern emerging for a given $\eta$ value,  we have performed simulations for the same variance but different distributions $W(\delta)$. The pattern depicted in Fig.~\ref{fig2}e refers to a setup with just two delays with equal weight such that
$M_2=1.33$ as in panel (d). They are quite similar to one another, suggesting that the variance is the (only) relevant
parameter of the delay distribution. This inference is confirmed by the equal shape of the autocorrelation function 
(see later for a quantitative analysis).
 
 A more stringent test of this empirical observation can be performed by studying a distribution such that $M_2=0$
 as in the single-delay case. This is indeed possible, since $W(\delta)$ can be negative for some
 $\delta$ value (although, obviously, $\int d \delta W(\delta) =1$ by definition).
 We have performed a test, by using the quasi-distribution
\begin{equation}
    W(\delta) = \frac{1}{q \sqrt{2\pi}} \exp\left( -\frac{\delta^2}{2 q^2} \right) \left[ 1 + \alpha \left( \frac{\delta^2}{q^2} - 1 \right) \right]~,
\label{quasi}
\end{equation}
suitably normalized and with zero mean. Its variance is
\begin{equation}
    M_2 =  q^2 (1 + 2\alpha)~.
\end{equation}
and can be negative if $\alpha$ is sufficiently negative.
The resulting pattern is reported in Fig.~\ref{fig2}c for $q = 1$ and  $\alpha = -0.5$ so that $M_2=0$. As seen, we have a strong agreement with the single delay case (Fig.~\ref{fig2}b) and even the correlation functions overlap.

For $\alpha <-0.5$, $M_2$ can even become negative. In order to test the behavior in one such case, we have reverted to the simple setup with two delays, choosing their values and strengths such that the $M_2=-0.32$. The resulting spatio-temporal representation is plotted in Fig.~\ref{fig2}a (see the caption for the details). Again, we find a pattern very similar to that for the single delay, but now the structures are {\it shrinked}: the autocorrelation peaks are 
indeed narrower here, keeping all the other features.

Summarizing the numerical findings, in the presence of a relatively narrow distribution of long-delay feedbacks,
the dynamics is very similar to that of a process characterized by a single delay equal to the mean value.
This is true also when the distribution is not trivial, as seen above. However, besides the pattern similarity,
it has emerged that the size of the coherent structures depends on the shape of the distribution $W(\delta)$: it can
either increase or decrease.

In the next section, we develop a theoretical approach which supports the numerical observations,
providing a perturbative expression for the dependence of the size of the spatial structures on the weight distribution $W(\delta)$.

\subsection{Theory}\label{sec:theory}
We start by presenting a linear stability analysis of the homogeneous solution $x=0$, to
show that the transition point does not change with respect to the single delay case. Let $u(t)$ denote an infinitesimal perturbation.
Given the presence of two (main) time scales, we assume
\begin{equation}
 u(t) = u_0 {\rm e}^{(\Lambda_0 + \Lambda_1/T)t}
\end{equation}
where $\Lambda_0=\lambda_0+i\omega_0$ and $\Lambda_1=\lambda_1+i\omega_1$ are complex numbers
 describing the stability over the two different time scales.
 Upon inserting this Ansatz into the linearized version of Eq.~(\ref{eq:model}), we obtain 
\begin{equation}
\Lambda_0 =  -1 + \eta {\rm e}^{-\Lambda_0 T-\Lambda_1}\langle {\rm e}^{-\Lambda_0 \delta_j} \rangle \; ,
\label{eq:linpert}
\end{equation}
where we have neglected $1/T$ terms.
Self-consistency requires $\lambda_0$ to be
equal to zero (this is a standard property of delayed models with a single variable, where all Lyapunov
exponents scale as $1/T$). Additionally, in the large $T$ limit, we are  authorized to assume 
$\omega_0 T \approx 2 \pi n$ for integer $n$.  As a result,
\begin{equation}
i\omega_0 =  -1 + \eta {\rm e}^{-\Lambda_1+K(\omega_0)}
\label{eq:linpert2}
\end{equation}
where 
\begin{equation}
K(\omega_0) =  \ln \langle {\rm e}^{-i \omega_0 \delta_j}\rangle  
\label{eq:linpert3}
\end{equation}
is the cumulant generating function. Notice that $K(\omega_0)$ represents the correction to the single delay case
 $\Lambda_1^s$ ($\Lambda_1 = \Lambda_1^s+K(\omega_0))$.
Expanding up to the second order in $\omega_0$, we formally find
\begin{equation}
K(\omega_0) \approx -i\omega_0 M_1 - \frac{\omega_0^2}{2}M_2 \;
\label{eq:dispers}
\end{equation}
where $M_1 =  \langle \delta_j \rangle = 0$ and $M_2 =  \langle \delta_j^2\rangle - \langle \delta_j\rangle^2$ are the first two 
(generalized) cumulants, so that 
\begin{equation}
\lambda_1 = \ln {\eta} - \frac{1}{2} \left( 1+ M_2 \right ) \omega_0^2
\label{lambda1}
\end{equation}
and
\begin{equation}
\omega_1 = -\omega_0 \;.
\end{equation}
From these equations, we see that $\omega_0=0$ is the most unstable wavelength and
the corresponding exponent, $\ln \eta$ becomes positive for $\eta>1$ exactly as in the single-delay case.
In other words, in a first approximation, we see that a model with multiple delays behaves as a single delay system, confirming the mean-field approach.

Having verified that a transition occurs for the same parameter value as for the single-delay case,  
we now proceed by formally introducing the above mentioned decomposition of the time axis $t$, 
by writing  write $x(t) = y(\sigma,\theta)$ and
$x(t-T_j) = y(\sigma-\delta_j,\theta-1)$.
In the vicinity of the bifurcation, it is convenient to write $\eta = 1+\varepsilon^2$, and
introduce the rescaled variables  $\xi=\varepsilon(\sigma-v\theta)$, $\tau=\varepsilon \theta$:
the introduction of a moving frame is dictated by a later simplification of the final equations if the velocity $v$ is properly selected.
By then following Ref.~\cite{Giacomelli1996}, we also expand the field in power series
\begin{equation}
y(\sigma,\theta) = \sum_{k=1}^\infty \varepsilon^k Z^{(k)}(\xi,\tau)~.
\end{equation}
This Ansatz has been substantiated by numerically verifying that the time-averaged field amplitude 
$A=\overline{X(t)}$ actually scales as $\varepsilon$ for different $M_2>0$ values.

Each feedback term can expanded as
\begin{equation}
x(t-T_j) =  \sum_{k=1}^\infty \varepsilon^k Z^{(k)}
\Big(\xi- \varepsilon(\delta_j-v) ,\tau- \varepsilon^2 \Big) \equiv \sum_{k=1}^\infty \varepsilon^k Z_j^{(k)}
\end{equation}
By averaging over the feedback channels and expanding $Z^{(k)}$ around $(\xi,\tau)$, 
under the assumption that  $M_1= 0$ we find
\begin{equation}
\langle Z_j^{(k)}\rangle=
Z^{(k)} +\varepsilon v Z_\xi^{(k)}+\varepsilon^2 \left(\frac{1}{2} (M_2+v^2) Z_{\xi\xi}^{(k)} -Z_\tau^{(k)}\right)+..
\end{equation}
where the subscript indicates a derivative with respect to that variable.
Analogously, the time derivative of the field can be expanded as
\begin{equation}
\dot x = \partial_\sigma y \to  \sum_{k=1}^\infty \varepsilon^{k+1} Z^{(k)}(\xi,\tau) \; .
\end{equation}
By finally including the expansion of the nonlinear term and equating the various orders, we see that 
the condition for the linear terms in $\varepsilon$ is automatically satisfied, while that for the
$\varepsilon^2$ terms is
\begin{equation}
Z^{(2)}= vZ^{(2)}
\end{equation}
which implies $v=1$ - the drift velocity is obtained.
At the third order, we finally obtain the relevant equation which describes the evolution
\begin{equation}
Z_\tau^{(1)} = Z^{(1)} + \frac{1}{2}(1+ M_2)Z_{\xi\xi}^{(1)} - (1+i\beta)|Z^{(1)}|^2 Z^{(1)}
\label{CGL}
\end{equation}
This CGL equation reduces to the known expression~\cite{Giacomelli1996} in the case of single delay, but is also
valid in the case of a zero-variance distribution as (\ref{quasi}), when higher moments are different from zero. 
Eq.~(\ref{CGL}) represents the main result of the theoretical part of this work.

At leading order, dispersion among feedback channels only modifies 
the diffusion coefficient $D$ (i.e. the coefficient in front of the second spatial derivative) of Eq.(\ref{CGL}). Since a variation of $D$ can be absorbed by rescaling the spatial axis, we can conclude that no qualitative changes of the overall dynamics are expected in the presence of multiple delays, as indeed visually seen while comparing
the panels of Fig.~\ref{fig2}. 

\begin{figure}
\includegraphics[width=1.0\linewidth]{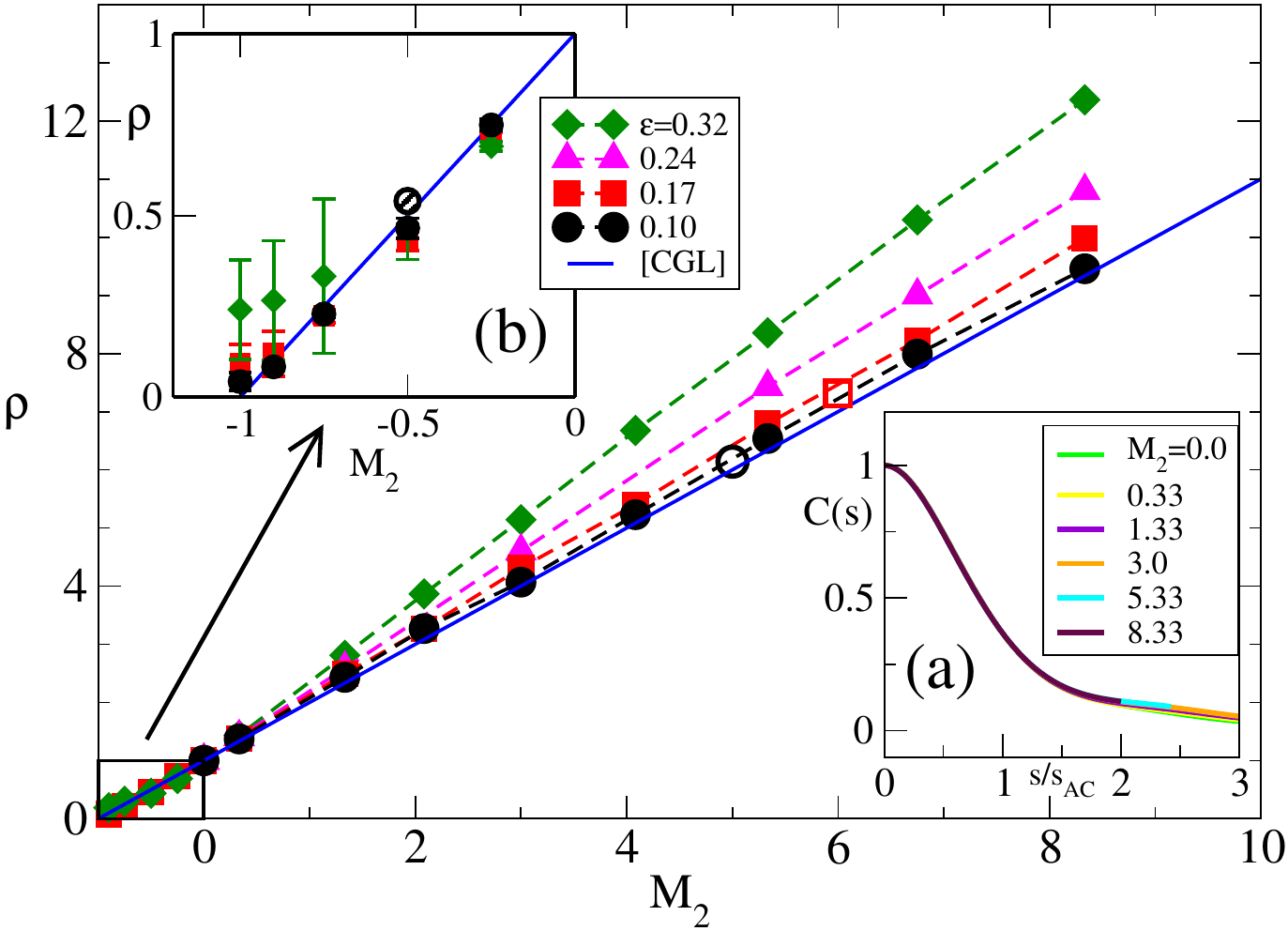}
\caption{Rescaled diffusion coefficient $\rho$ for different values of $\varepsilon$ (see legend) at $T=4096$; the dashed lines are guides for the eye, and the blue solid line is the scaling predicted by (\ref{CGL}). Inset (a): normalized correlations for $\varepsilon=0.17$, at the $M_2$ values in the legend, rescaled by the correlation length as defined in the text. Inset (b): enlargement of the negative regimes of $M_2$ in the box. The empty symbols are obtained from a two-delay distribution (see Tab. I for the $M_2>0$ case and Tab. II for the $M_2<0$ case). Other simulations are for a flat delay distribution, and the barred circle in (b) corresponds to distribution (\ref{quasi}) with $q = 1, \alpha = -0.75$. }
\label{fig3}
\end{figure}

On a more quantitative level, Eq.~(\ref{CGL}) implies that the spatial scales change by a factor $\sqrt{1+M_2}$. Introducing the correlation length $s_{AC}$ as the ``spatial" distance such that $C(s_{AC}) = {\rm e}^{-1}$, we expect that
\begin{equation}
\rho(M_2) =  \left( \frac{s_{AC}(\varepsilon,M_2)}{s_{AC}(\varepsilon,0)}\right)^2 \approx 1 + M_2
\end{equation}
This prediction is thoroughly investigated in Fig.~\ref{fig3}, where the results obtained for different distributions, with several $M_2$ and 
$\varepsilon$ values are reported (recall that $\rho(0)=1$ by definition). 

The different symbols refer to different $\varepsilon$ values. The solid blue straight line is the theoretical prediction. We start noticing that 
in all cases, $\rho(M_2)$ increases linearly with $M_2$, For $\varepsilon=0.10$ (the smallest value herein considered) the slope is in a fairly good agreement with the theoretical prediction (see the full black dots). Upon increasing $\varepsilon$, the slope increases a bit, suggesting that higher order terms (here not considered) are no longer fully negligible. We also performed some simulations for the two delays distribution to validate the prediction that
only the value of $M_2$ matters (see the parameters in Tab. I and the empty points in Fig.~\ref{fig3}). 

\begin{table}[htbp]
\caption{\label{tab+}Parameters and results for the examples in the case of two-delays distribution, for $M_2>0$.}

\begin{tabular}{|c|c|c|c|c|c|c|c|}
\hline 
$\varepsilon$ & $M_2$ & $w_1$ & $\delta_1$ & $w_2$ & $\delta_2$ & $\rho$ &1+$M_2$\\ 
\hline 
0.10 & 5.0 & 0.8333 & -1.0 & 0.1667 & 5.0 & 6.14 & 6.0 \\ 

\hline 
0.17 & 6.0 & 0.8572 & -1.0 & 0.1428 & 6.0 & 7.31 & 7.0 \\ 

\hline 
\end{tabular} 

\end{table}

As already mentioned, $M_2$ can also assume negative values;  we have generated them by considering
suitable two-delay distributions (see Tab. II for the parameters). 
The corresponding results can be seen in the lower left corner of Fig.~\ref{fig3} and in fact they again confirm the theoretical prediction.
The same data are presented also in the upper-left inset, where one can also appreciate differences among different realizations of the same variance $M_2$:
they are quantified by the the error bars,  which represent the standard deviation of the fluctuations among three different ways of obtaining a given $M_2$ value
(see Tab II, for quantitative details). In practice, we see that for larger $\varepsilon$ values (around 0.3), the shape of the weight distribution matters, indicating
that a more refined theory becomes  necessary. This sensitivity is a manifestation of the vicinity to the critical point $M_2=-1$,  below which
the overall diffusion coefficient of the corresponding CGL equation becomes negative and cannot any longer reproduce the pattern generated by the delayed model. 
Mathematically, from  Eq.~(\ref{lambda1}), we see that $\omega_0$ transforms itself from a maximum into a minimum of the stability spectrum, 
implying that additional terms must be included to ensure stability, i.e. higher-order spatial derivatives need be added to the CGL equation or, equivalently,
higher-order terms should accounted for in the multiscale expansion to ensure even a qualitative agreement.
Nonetheless, we see in Fig.~\ref{fig2}a that the pattern generated for a negative $M_2=-0.32>-1$ contains the same typeof coherent structures as in the single-delay case, just
with narrower triangles.

\begin{table}[htbp]
\caption{\label{tab-}Parameters for the two-delays distributions with $M_2<0$.}

\begin{tabular}{|c|c|c|c|c|}
\hline 
$M_2$ & $w_1$ & $\delta_1$ & $w_2$ & $\delta_2$\\ 
\hline 
-0.25 & -1/35 & 3 & 36/35 & 1/12 \\ 

\hline 
-0.50 & -1/17 & 3 & 18/17 & 1/6 \\ 

\hline 
-0.75 & -1/11 & 3 & 12/11 & 1/4 \\ 

\hline 
-0.75 & -3/13 & 2 & 16/13 & 3/8 \\ 

\hline 
-0.75 & -1/2 & 3/2 & 3/2 & 1/2 \\ 

\hline 
-0.90 & -1/9 & 3 & 10/9 & 3/10 \\ 

\hline 
-0.90 & -9/31 & 2 & 40/31 & 9/20 \\ 

\hline 
-0.90 & -2/3 & 3/2 & 5/3 & 3/5 \\ 

\hline 
-1.00 & -1/8 & 3 & 9/8 & 1/3 \\ 

\hline 
-1.00 & -1/3 & 2 & 4/3 & 1/2 \\ 

\hline 
-1.00 & -4/5 & 3/2 & 9/5 & 2/3 \\

\hline 
\end{tabular} 

\end{table}

In order to further confirm the correctness of the theoretical predictions, we have investigated the entire profile of the
autocorrelation function to verify it is independent of the delay, once properly rescaled. Various curves, obtained for the
same $\varepsilon=0.17$ and different values of $M_2$ are reported in the lower-right inset of Fig.~\ref{fig3}, after a rescaling
of the spatial axis. They show a  good overlap over a broad range of values thus further confirming that the dynamics altogether
is unchanged.

As a last test, we consider the correlation length $\theta_{AC}$, expressed in number of delay units such that the height of the recurrent peaks, visible
in Fig.~\ref{fig1} decays by  a factor $1/{\rm e}$. Our perturbative theory predicts that the time scale of the spatio-temporal dynamics is invariant in the presence of multiple delays. This can be qualitatively appreciated by inspecting the patterns in Fig.~\ref{fig2}; a quantitative numerical analysis for a flat distribution and  $\varepsilon = 0.10$ shows that the correlation length along $\theta$  passes from $\theta_{AC}=245.2$ for the single delay (where $M_2=0$), to $\theta_{AC}=248.9$ for $M_2=1.33$ and to $\theta_{AC}=253.2$ for $M_2=5.33$. In other words, the
variation is very small (0.4$\%$ increase vs $255 \%$ for the spatial scale), substantially confirming the predicted independence of $M_2$.

\subsection{Conclusions}\label{sec:conclusion}
Altogether, we have shown  that in the presence of multiple delays, so long as the time scale of the dispersion is of the same order as that of
the local dynamics, the overall behavior can be still described in terms of an equivalent PDE. Interestingly,
we find that adjusting the coupling strengths, the resulting amplitude of the diffusion coefficient may be tuned in both directions, i.e. it can be increased as well as decreased.
We have limited ourselves to considering  a strictly real feedback, but there are no restrictions to extending the method to
complex coupling with rotations, which would be very appropriate especially in the context of networks of optical fibers~\cite{LANER,Tassi2026}.

As seen, while for deviations from the bifurcation of order $\varepsilon \approx 0.1$, the theoretical description is quantitatively accurate,
larger $\varepsilon$ values lead to discrepancies which indicate the need to incorporate higher-order terms. Preliminary studies suggest that
many additional terms have to be included. We plan to further investigate the question in the hope of obtaining a relatively simple effective description.

Finally, we remark that the dynamics depends only on the variance $M_2$ of the delay distribution. Thus, the system is {\it resilient}: especially in the presence of a large number of feedback terms, fluctuations of couplings and delays have a very low impact. Even more so, considering that the dependence on
the variance consists essentially only in an adjustment of the pseudo-spatial coherent structures.

\subsection{Acknowledgments}
G.G. acknowledges support from the Research Project PRIN 2022 "The structure, dynamics and control of network systems with higher-order interactions", funded by the Italian Ministry of University and Research (n. 2022FHHHPC).

\end{document}